\documentclass[twocolumn]{aastex631}

\usepackage{multirow}
\usepackage{array}
\usepackage{tabularx}

\usepackage{newtxtext,newtxmath}

\usepackage[T1]{fontenc}
\usepackage[flushleft]{threeparttable}

\DeclareRobustCommand{\VAN}[3]{#2}
\let\VANthebibliography\thebibliography
\def\thebibliography{\DeclareRobustCommand{\VAN}[3]{##3}\VANthebibliography}

\usepackage{epsfig}
\usepackage{epstopdf}
\usepackage{amsmath}
\usepackage{multirow}

\newcommand{\mum}{\textmu m}
\newcommand{\q}{`}
\newcommand{\sfr}{\,M$_{\odot}$yr$^{-1}$}
\newcommand{\sol}{\,M$_{\odot}$}

\accepted{August 30, 2025}

\shorttitle{A merger-driven massive galaxy at $z = 3.65$}
\shortauthors{Pitchford et al.}

\begin{document}

\title{Witnessing the violent, merger-driven formation of an extremely massive galaxy\\ 1.7\,Gyr after the Big Bang}

\correspondingauthor{L.K. Pitchford}
\email{lura.k.pitchford@gmail.com}

\author[0000-0002-5206-5880]{L.K. Pitchford}
\affiliation{Department of Physics and Astronomy, Texas A\&M University, College Station, TX 77843-4242, USA}
\affiliation{George P. and Cynthia Woods Mitchell Institute for Fundamental Physics and Astronomy, Texas A\&M University, College Station, TX 77843-4242, USA}

\author{J. Cairns}
\affiliation{Astrophysics Group, Imperial College London, Blackett Laboratory, Prince Consort Road, London SW7 2AZ, UK}

\author[0000-0003-1748-2010]{D. Farrah}
\affiliation{Department of Physics and Astronomy, University of Hawaii, 2505 Correa Road, Honolulu, HI 96822, USA}
\affiliation{Institute for Astronomy, 2680 Woodlawn Drive, University of Hawaii, Honolulu, HI 96822, USA}

\author[0000-0002-9548-5033]{D.L. Clements}
\affiliation{Astrophysics Group, Imperial College London, Blackett Laboratory, Prince Consort Road, London SW7 2AZ, UK}

\author[0000-0003-0917-9636]{E. Hatziminaoglou}
\affiliation{ESO, Karl-Schwarzschild-Str. 2, 85748 Garching bei M\"unchen, Germany}
\affiliation{Instituto de Astrofísica de Canarias, C/Vía Láctea, s/n, E-38205 San Cristóbal de La Laguna, Tenerife, Spain}
\affiliation{Universidad de La Laguna, Dpto. Astrofísica, E-38206 San Cristóbal de La Laguna, Tenerife, Spain}

\author[0000-0002-2807-6459]{I. Pérez-Fournon}
\affiliation{Instituto de Astrofísica de Canarias, C/Vía Láctea, s/n, E-38205 San Cristóbal de La Laguna, Tenerife, Spain}
\affiliation{Universidad de La Laguna, Dpto. Astrofísica, E-38206 San Cristóbal de La Laguna, Tenerife, Spain}

\author[0000-0002-6736-9158]{L. Wang}
\affiliation{SRON Netherlands Institute for Space Research, Landleven 12, 9747 AD, Groningen, The Netherlands}
\affiliation{Kapteyn Astronomical Institute, University of Groningen, Postbus 800, 9700 AV Groningen, the Netherlands}


\begin{abstract}

We combine near-infrared imaging in two bands from the Hubble Space Telescope (HST) with archival observations of molecular gas to study SDSS J160705.16+533558.6 (J1607), an extremely luminous broad-line quasar at $z = 3.65$ that is also bright in the submillimeter (sub-mm). Via subtraction of the quasar point spread function, we show that its host galaxy is massive, with a stellar mass of $(5.8 \pm 3.0) \times 10^{11}$\sol, making it comparable to giant early-type galaxies (ETGs) at $z\sim0$. 
If the supermassive black hole (SMBH) in the quasar is accreting at the Eddington limit, then its mass is $3.5 \times 10^{9}$\sol, which is also consistent with local massive ETGs. 
The host has an extremely high star formation rate (SFR) of $4300 \pm 500$\sfr and a molecular gas mass of $(2.4 \pm 0.9)\times 10^{10}$\sol. 
The quasar has two companions: one at a projected separation of 11\,kpc with a stellar mass of $(7.9 \pm 5.0) \times 10^{10}$\,M$_{\odot}$ but no detected molecular gas, and one 6\,kpc further away in the same direction with a molecular gas mass of $(2.6 \pm 1.3) \times 10^{10}$\,M$_{\odot}$ but no detected stellar emission.
Since neither companion shows evidence for AGN activity, this may represent merger-driven quenching, in which the dynamics of the merger strip molecular gas from infalling galaxies. 
Overall, irrespective of whether the host is merging with the companions, these properties mark J1607 as forming what will become an extremely massive ($\sim10^{12}M_{\odot}$) galaxy by $z=0$.

\end{abstract}




\section{Introduction} \label{sec:intro}

Among the most striking results from Herschel was the discovery of a population (2--3 per square degree) of extremely luminous star-forming galaxies, with star formation rates (SFRs) exceeding about 2000\sfr, at $z\gtrsim2$  \citep{dowell14,asboth16,gao21, wang21}. 
Examples include SPT0346-52 at $z=5.66$ \citep[forming stars at a rate of 4500\sfr;][]{ma16}, HFLS3 and SPT0311–58 at $z=6.34$ and $z=6.90$ \citep[both at 2900\sfr;][]{riechers13,marrone18}, ADFS-27 at $z=5.66$ \citep[$2400$\sfr;][]{riechers17}, HXMM01 at $z=2.31$ \citep[$2000$\sfr;][]{fu13}, and NA.v1.489 at $z=2.69$ \citep[$1900$\sfr;][]{nayy17}.  
However, this population has proved challenging to study. 
There are no direct analogs at low redshifts, since virtually all sources at $z<1$ have SFRs at least an order of magnitude lower than those seen in the extreme starburst population.  
Galaxy formation models either do not contain them at all, or do so at space densities an order of magnitude or more below what is observed  \citep{baugh05, lacey10, gruppioni11, guo11, hayward13}. 
The trigger mechanisms and evolutionary pathways for extreme starburst galaxies thus remain unclear. 
 
Although they are challenging to study, extreme starbursts are suspected to play an important role in the formation of massive quiescent galaxies. 
Such quiescent systems are already observed by $z \sim 2$ \citep[e.g.][]{szomoru12, whitaker12, toft14}, making extreme starbursts plausible antecedents for them \citep{wellons15, dave17}. 
There is also evidence of a tight spread in stellar ages in (some) quiescent galaxies at $z \sim 2$, meaning that all of the stars in these galaxies must have formed at roughly the same time \citep[e.g][]{glazebrook17}.
The formation histories of low-redshift massive quiescent galaxies are however complex and not yet fully understood \citep[e.g.][]{mcdermid15,farrah23}, so their links with high-redshift extreme starbursts have yet to be clarified. 
 
Some insights into extreme starbursts can be gained from the most actively star-forming galaxies at low redshift. 
These galaxies, the ultraluminous infrared galaxies (ULIRGs; galaxies with infrared luminosities above $10^{12}$L$_\odot$), typically have SFRs a factor of 4-20 times lower than those seen in high-redshift extreme starbursts.
ULIRGs at $z\lesssim0.2$ are almost always mergers \citep[e.g.][]{clements96,farrah01}. 
At higher redshifts, some ULIRGs are found in merging systems \citep[e.g.][]{zamojski11,ivison12}, while others are not \citep{melbourne09, wiklind14}.
The merger fraction among ULIRGs likely declines with increasing redshift  \citep[e.g.][]{kartaltepe12}, though it is not known if this also applies to the most luminous starbursts.  
ULIRGs at $z\lesssim0.2$ virtually always harbor active star formation, and in about half of them, the star formation is accompanied by a luminous AGN \citep{efstathiou22,farrah22}.  
At higher redshifts, the SFRs associated with ULIRGs are higher \citep[e.g.][]{michalo17,malek18,gullberg18}, and many are again accompanied by an AGN \citep{farrah16,farrah17}. 
A small number of quasars at $z\gtrsim1$ with SFRs exceeding 2000\sfr are known \citep[e.g.][]{pitchford16}, and rapidly star-forming galaxies adjacent to quasars at very high redshift have been identified \citep{decarli17}.  
However, to date, case studies of extreme starbursts at high redshift have focused on systems for which there is no evidence for a luminous AGN \citep{fu13,riechers13,ma16,riechers17,nayy17,marrone18,pavesi18}. 
Only a few well-studied extreme starbursts also harbor AGN \citep[e.g.][]{shao19}.

Case studies of high-redshift extreme starbursts that also host luminous AGN are thus of value in helping place this population within the context of galaxy assembly.  
We here present such a study of SDSS J160705.16+533558.6 (hereafter J1607), a system previously shown to harbor both a broad-line luminous quasar with L$_\textnormal{AGN} \sim 10^{14}$\,L$_\odot$ and a starburst with L$_\textnormal{SB} \sim 10^{13.5}$\,L$_\odot$ \citep{clements09} at a redshift of 3.65 \citep{adelman08}. 
Submillimeter (sub-mm) observations further show a candidate companion, an $\sim$10\,kpc emission region that is offset from the quasar by about 1.5\arcsec ($\sim$11\,kpc) to the northwest. 
This companion is consistent with a merger \citep{clements09}. 
To describe the components of this system (quasar, host, and potential companion) and determine what role, if any, the $z > 3$ luminous starburst population might play in assembling local ellipticals, we have obtained Wide Field Camera 3 (WFC3) imaging from the Hubble Space Telescope (HST) in F110W and F160W.  
Extracting the host requires removing the quasar through subtracting the point spread function (PSF).  
To effectively remove the quasar, we create two types of PSFs, empirical and synthetic, and compare the resultant subtractions.  
These PSF-subtracted images allow us to better describe the host galaxy and its companion.  
We combine the galaxies' properties with archival data to set the whole system within the broader context of how obscured luminous starbursts are triggered and evolve. 

This paper is structured as follows. 
In \S \ref{sec:hst_observations}, we introduce J1607 and describe the HST observations. 
In \S \ref{sec:psf_creation} and \S \ref{sec:psf_subtraction}, we describe the PSF creation and subtraction methods, respectively. 
We make use of two different sets of PSFs, the empirical PSFs being built from a nearby star and the synthetic PSFs being created in \textsc{Tiny Tim}. 
We then present our results in \S \ref{sec:results}, place them into context with other galaxies in \S\ref{sec:discussion}, and present our conclusions in \S \ref{sec:conclusions}. 
Throughout this work, we assume $H_0 = 70$\,km\,s$^{-1}$Mpc$^{-1}$, $\Omega_\Lambda = 0.7$ and $\Omega_\textnormal{M}=0.3$.


\section{HST Observations}\label{sec:hst_observations}

Given its high redshift, we observed J1607 using the HST WFC3 IR channel, which has a pixel scale of 0.13\arcsec, with two orbits each of the F110W (Wide \textit{YJ}) and F160W (WFC3 \textit{H}) filters as part of Program ID 15200 on 2018 January 28 and 2018 February 10. 
We used a four-point \textsc{box-min} dither pattern with the default point and line spacings of 0.572\arcsec and 0.365\arcsec, respectively. 
We also used SPARSE sampling so that the nucleus of the quasar could not saturate. 
Because the sub-mm companion is located more than 0.5\arcsec\,from the quasar, we did not request a separate PSF observation, but instead build a PSF from a nearby star as described in \S\ref{sec:image_psf_creation}. 
Each dither had an exposure time of 702.9\,s for a total on-source time of 2811.8\,s per orbit. 
We then use \textsc{drizzlepac} \citep{gonzaga12} to combine each individual calibrated (\textit{\_flt}) exposure.
The HST observations were obtained from the Mikulski Archive for Space Telescopes (MAST) at the Space Telescope Science Institute (STScI) and can be accessed via \dataset[doi:10.17909/c8wq-7923]{https://doi.org/10.17909/c8wq-7923}.


\section{PSF Creation}\label{sec:psf_creation}

To extract the flux of the star-forming companion and search for the underlying host, we must first subtract the quasar.
Similar PSF subtraction methods have been applied to other high-redshift systems in the past, with varying levels of success \citep[see e.g.][]{guo09, mechtley12, mechtley16, marian19, marian20, marshall20}.
We subtract the PSF using two methods, which differ only in how the PSFs are created. 
We refer to these PSFs as the empirical and synthetic PSFs. 
Because diffraction spikes are artifacts caused by the structure of the telescope, the empirical PSFs are likely to better model them than are the synthetic PSFs. 
As such, we expect the empirical PSF subtractions to provide more accurate subtractions of these features. 
We therefore rely on the empirical PSFs even though the synthetic PSFs are more widely used in the literature \citep[e.g.][]{vandokkum08, vanderwel11, villforth14} as they should provide a better overall subtraction \citep[see also e.g.][]{mechtley16}.  

Comparing the empirical PSFs (and the resultant subtractions) to the synthetic also allows us to test the standard approach, i.e. \textsc{Tiny Tim}, against something that is more accurate, the empirical approach, and shows that an empirical approach can be applied even in the absence of numerous stars in the field. 
While some studies require multiple stars distributed over the whole field-of-view to accurately sample the PSF \citep[e.g.][]{glikman15}, we find that a single star within about 20\arcsec of the target provides an excellent sampling of the effective HST optics and WFC3 detector.
Once we have our best subtraction, we compare the results to those from a more complex fitting algorithm, \texttt{GALFIT} \citep{peng11}\footnote{\url{http://users.obs.carnegiescience.edu/peng/work/galfit/galfit.html}}.

\subsection{Empirical PSF Creation}\label{sec:image_psf_creation}

\begin{figure}
    \centering
    \includegraphics[width=0.49\linewidth]{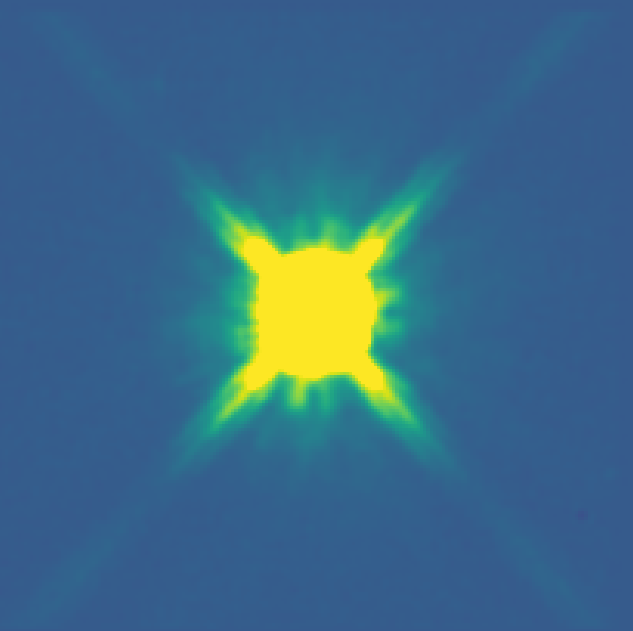}
    \includegraphics[width=0.49\linewidth]{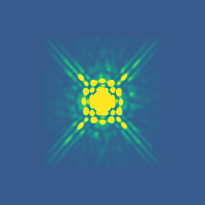}
    \caption{Empirical (left) and synthetic (right) PSFs for F110W. Each image is $51 \times 51$\,pixels ($\sim$6.6\arcsec $\times$ 6.6\arcsec). As the empirical PSF is created using a star in the same field as J1607, it better describes the telescope optics at the time of observation than does the synthetic PSF.}
    \label{fig:psf_comp}
\end{figure}

The empirical PSFs are derived using the Python \textsc{photutils} package \citep{bradley21}, which builds the effective PSF first described by \citet{anderson00}. 
We derive an effective PSF for each of our orbits from the background-subtracted images using a single, unsaturated star in our field that is relatively close to the quasar (within 20\arcsec).
The NIR (F110W - F160W) color of our PSF star differs from that of our quasar by 0.1 mag. 
We are mostly concerned with removing the diffraction spikes as one passes through the companion, so we do not expect the slight color mismatch to affect our results.
We set the size of each PSF to be $51 \times 51$\,pixels ($\sim$6.6\arcsec $\times$ 6.6\arcsec) in the native scale in order to include the diffraction spikes and to have the brightest pixel be the central pixel. 
Because our empirical PSFs should be unaffected by e.g. spatial variations across the field or time variations due to the instrument focus, we expect the empirical PSFs to better describe (and later remove) the diffraction spikes when compared against the synthetic PSFs (see \S\ref{sec:tinytim_psf_creation}). 
The normalization radius is set to four pixels to enclose the central region of the quasar, but not the star-forming companion, in order to scale the PSF to match the flux of the quasar. 
We then oversample by a factor of four to make sub-pixel shifts in the position of the PSFs relative to each of the calibrated images in order to improve the accuracy of the final PSF subtraction. 
This is done because the WFC3 native scale is significantly undersampled. 
The use of the four-point dither for our observations does mitigate this, but the allowance of sub-pixel shifts provides an even more accurate PSF; an oversampling of four is chosen as \citet{anderson16} finds it to work well for WFC3/IR images. 
To ensure that the brightest pixel is the central pixel, each axis is made odd, i.e. the oversampled PSF is $205 \times 205$\,pixels in size. 
The empirical PSF is scaled such that the sum of its pixels is the square of the oversampling factor. 
One of the empirical PSFs is provided in the left panel of Fig. \ref{fig:psf_comp}.

\subsection{Synthetic PSF Creation}\label{sec:tinytim_psf_creation}

The synthetic PSF models are generated using the STScI program \textsc{Tiny Tim}, as it has long been the standard synthetic PSF-modelling software for HST observers. 
\textsc{Tiny Tim} is split into three programs. 
\textsc{tiny1} requests a series of inputs from which to generate a PSF model, including the coordinates of the source on the detector grid, the filter used for the observations, the diameter of the PSF in arcseconds, and the desired oversampling of the PSF model. 
\textsc{tiny2} then uses these inputs to generate a slightly oversampled and undistorted model PSF. 
The WFC3 and ACS instruments do suffer from significant distortion that is not corrected for by the internal optics \citep{krist11}, however, as they are located away from the optical axis of the HST instrument. 
Therefore, \textsc{tiny3} applies a geometric distortion to the PSF generated by \textsc{tiny2} to account for this effect.

We generate a synthetic PSF model for each filter centered on the rough position of the quasar on the detector array, with a diameter of 4\arcsec. 
Additionally, we follow the method of \cite{biretta12} and \cite{biretta14} and edit the \textsc{Tiny Tim} Cold Mask parameters to their improved values to better describe artifacts such as diffraction spikes. 
We generate the final synthetic PSFs oversampled by a factor of four to match that of the empirical PSFs. 
Each model PSF produced by \textsc{Tiny Tim} is scaled such that the sum of its pixels is one. One of the synthetic PSFs is shown in the right panel of Fig. \ref{fig:psf_comp}.


\section{PSF Subtraction}\label{sec:psf_subtraction}

Once we have the empirical and synthetic PSFs, we subtract them from each individual calibrated image (corresponding to each dither position) prior to drizzling them together. 
We employ the same basic method for the subtractions. 
Once the PSF fluxes are scaled to match the quasar flux, we oversample the full field by a factor of four, the same oversampling factor used to create the PSFs. 
To help determine the best alignments between the PSFs and the quasar images, we manually inspect a grid of oversampled quasar cutouts with the same size as the PSFs and with centers in a $10 \times 10$\,pixel box around the oversampled position corresponding to the native scale center. 
Once the best visual alignment between the PSF and the oversampled quasar cutout is found, we subtract the two.
We then paste the PSF-subtracted cutout back into the original image and regrid the image back into the native pixel scale of the observations. 
We repeat this for each dither for both orbits before drizzling together the eight frames to make a single image for each of the filters.

\subsection{Empirical PSF Subtraction}\label{sec:image_psf_sub}

The empirical PSFs are normalized to the square of the oversampling factor, so we must scale the PSF using the ratio of the quasar flux to the unscaled PSF flux. 
This results in the PSF flux densities given in the first and fourth rows of the first column of Tab. \ref{tab:sub_comp}. 
Errors are estimated using the differences in flux density from each dithered exposure. 

\begin{figure*}
    \centering
    \includegraphics[width=0.9\linewidth]{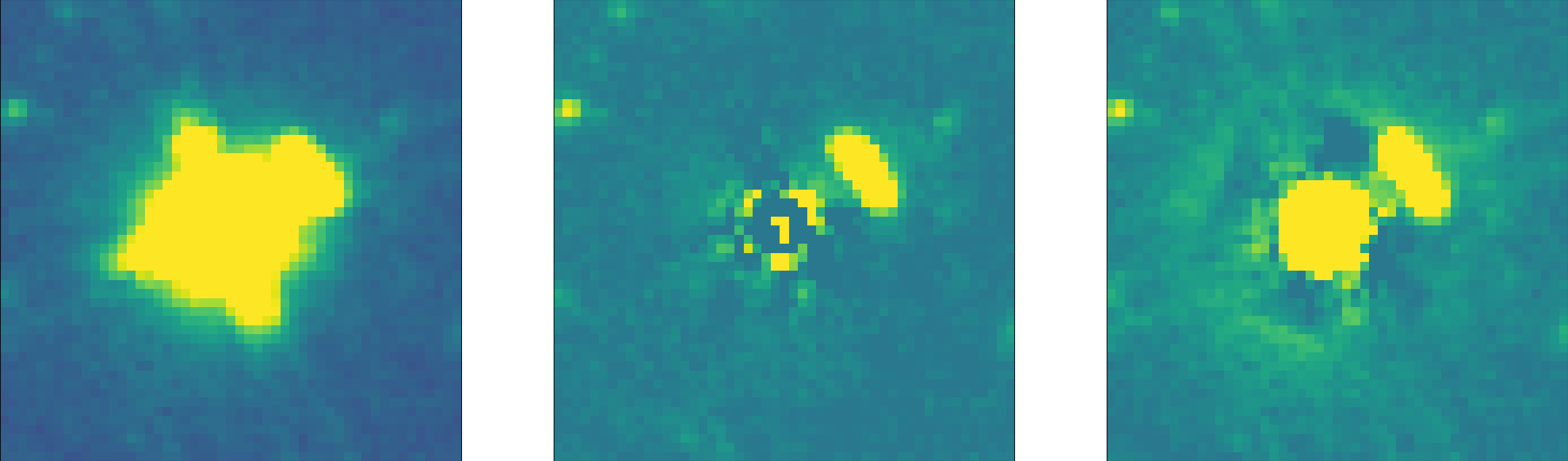}
    \includegraphics[width=0.9\linewidth]{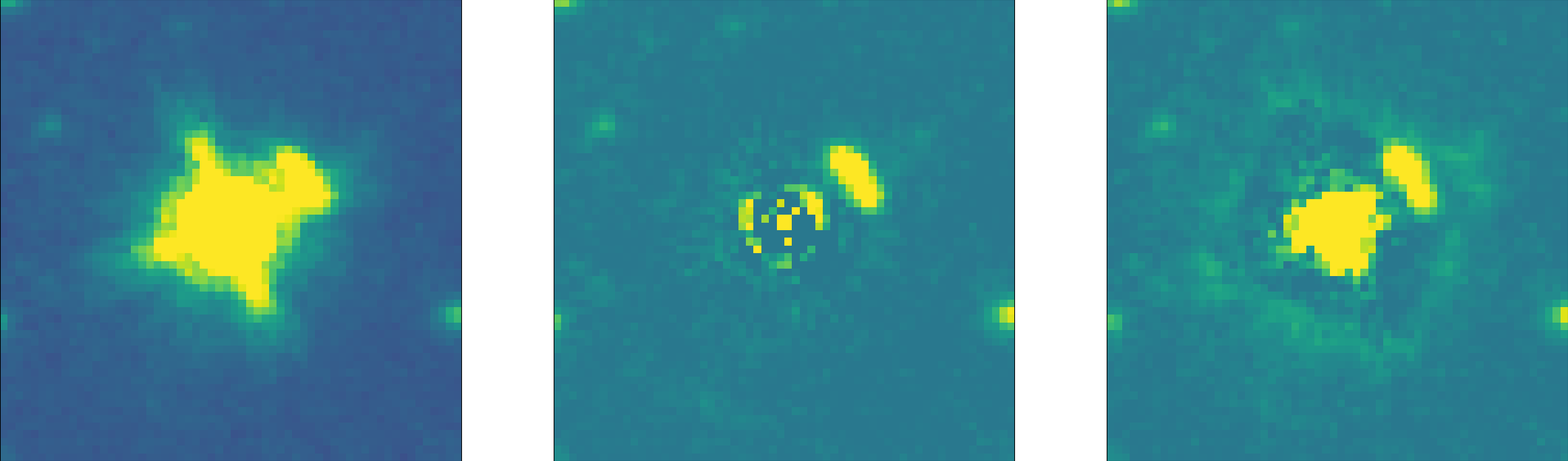}
    \caption{PSF subtraction results comparison. Each cutout is $60\times 60$ pixels (7.8\arcsec x 7.8\arcsec). The left column shows the unsubtracted images. The subtracted images using the empirical PSFs and the synthetic PSFs are shown in the middle and right columns, respectively. The top row shows the F110W images, while those for F160W are in the bottom row. In each image, north is up, and east is to the left. The scaling has been chosen to emphasize the companion galaxy; the two sets of subtracted images have a different scaling to the unsubtracted images. The empirical PSFs do a better job of removing the diffraction spikes.}
    \label{fig:psf_sub_comp}
\end{figure*}

We use the diffraction spikes to ensure the proper alignment for each dither. 
The best drizzled subtractions for each filter are shown in the middle column of Fig. \ref{fig:psf_sub_comp}. 
When compared to the unsubtracted drizzle products (the left column of the same figure), we see that the empirical PSFs are reasonably successful in removing the diffraction spikes.

\subsection{Synthetic PSF Subtraction}\label{sec:tinytim_psf_sub}

As the synthetic PSF is already scaled such that the sum of all pixel values is one, we can simply multiply the PSF by the quasar flux. 
Doing so gives the flux densities in the first and fourth rows of the second column of Tab. \ref{tab:sub_comp}. 

In subtracting the synthetic PSF, however, we cannot rely as heavily on the diffraction spikes to aid us in the alignment, as the synthetic PSFs are unable to accurately reproduce these features. 
Thus, we must also use the central quasar region to help align the PSFs to the images.
Using the unsubtracted images, we find that the Airy ring peaks at about 5\,pixels from the central pixel. 
We allow this ring to have a width of 2\,pixels, and use it in combination with the diffraction spikes to align the PSFs. 
The results from the synthetic PSF subtractions for each filter are shown in the right column of Fig. \ref{fig:psf_sub_comp}.
It is clear when comparing these to the empirical-PSF-subtracted images that the synthetic models are unable to account for the diffraction spikes. 
When considering the residuals, those from the empirical subtraction are smaller by more than a factor of three.
Because the synthetic models do not recreate the PSF as accurately as do our empirical models, we use the synthetic PSF subtractions instead to provide upper limits. 
We do also note the existence of empirical WFC3/IR PSFs created by STScI. 
However, because the user is unable to define the size of the PSF, these PSFs present similar issues to the synthetic PSFs in that we cannot fully remove the diffraction spikes from the image. 
Further, as the diffraction spikes change based on the telescope optics at the time of the observations, our empirical PSFs still most accurately define these features. 

\begin{deluxetable}{cccc}
    \tablecaption{Flux Density Comparisons for the Empirical and Synthetic PSF Subtractions}
    \label{tab:sub_comp}
    \tablehead{
        \colhead{HST Band} & \colhead{Component} & \colhead{Empirical Flux} & \colhead{Synthetic Flux}\\[-0.25cm]
        \colhead{} & \colhead{} & \colhead{Density} & \colhead{Density}\\[-0.25cm]
        \colhead{} & \colhead{} & \colhead{($\rm \mu Jy$)} & \colhead{($\rm \mu Jy$)}
    }
    \startdata
        {} & PSF & $241.3 \pm 7.2$ & $276.9 \pm 8.4$\\
        F110W & Host & $10.9 \pm 9.9$ & $35.7 \pm 12.6$ \\
        {} & Companion & $3.1 \pm 0.1$ & $3.3 \pm 0.1$\\
        \hline
        {} & PSF & $280.5 \pm 8.4$ & $320.3 \pm 7.2$ \\
        F160W & Host & $16.7 \pm 12.4$ & $51.7 \pm 14.9$\\
        {} & Companion & $3.3 \pm 0.1$ & $3.3 \pm 0.1$\\
    \enddata
    \tablecomments{The empirical and synthetic flux densities denote which PSF was used in the subtraction.}
\end{deluxetable}

\subsection{GALFIT PSF Subtraction}\label{sec:galfit_psf_sub}

As a final comparison, we run GALFIT with our empirical PSFs, as it is a more sophisticated algorithm than our subtraction method.
However, we try to match our initial method as much as possible. 
We therefore set up GALFIT in the following way.
We continue to work on the individual dithers, but provide a cutout of the region immediately surrounding the quasar to be fit. 
As we have subtracted the background already, we do not allow GALFIT to fit its own background (i.e. we input our background-subtracted images, not just the calibrated images from the HST pipeline). 
To ensure that the central pixels alone do not dictate the fit, we mask the central 3x3 pixels, as well as any bright pixels outside of the central region that GALFIT might otherwise try to fit. 

We fit three separate components in GALFIT: one PSF and two Sersic components (the quasar host and optical companion).
The cutout is centered on the quasar, which then also defines the center starting point for both the PSF and the host components.
The starting PSF magnitude is determined based on the total system magnitude.
Since the quasar outshines its host, the host magnitude starting point is an order of magnitude fainter.   
We further set constraints on both the effective radius and the Sersic index, with the former taking on values between 1 - 6 pixels and the latter taking on values between 1 and 5.
For the companion, the starting center, magnitude, axis ratio, and position angle are based on a combination of the original, unsubtracted images (for the center, axis ratio, and position angle) and the remnant companion in our initial subtractions (for the magnitude). 
They are all, however, still allowed to vary.


\section{Results}\label{sec:results}

\subsection{HST Photometry}\label{sec:hst_phot}

After subtracting the PSF, we extract the photometry of the host and the companion. 
As mentioned above, the Airy ring peaks at $\sim$5\,pixels and is taken to have a width of 2\,pixels. 
Therefore, to determine the flux density of the host, we use a circular aperture of radius 4\,pixels, and the results are given in the second and fifth rows of Tab. \ref{tab:sub_comp} for the two sets of PSF subtractions\footnote{We did also allow the host to encompass the full central region (radius of 6\,pixels), but the flux densities changed by at most five percent and thus the decision to exclude the Airy ring does not significantly change our results.}. 
We use an elliptical aperture centered on the companion to obtain its flux density for the two sets of subtractions; these are given in the third and sixth rows of Tab. \ref{tab:sub_comp}.

There are two main sources of error in our flux densities: the calibration error, which is $\sim$1$\%$ for the wide-band WFC3 filters \citep{mack21}, and the error that we introduce in completing the PSF subtraction. 
We find the former to dominate for the companion and the latter to dominate for the host. 
To obtain an estimate for the subtraction error that we are introducing, we allow the centers to change from the best values (i.e. the centers that best subtract the diffraction spikes). 
We take the standard deviation of those flux densities where the diffraction spikes appear aligned upon visual inspection to then be the subtraction-induced error. 
For the host, we additionally include the uncertainties on the PSFs themselves.
We do not have a strong detection of the host in either filter.

While the companion flux densities are relatively consistent, the host flux densities from the synthetic subtractions are higher than those from either of the empirical subtractions. 
GALFIT gives the lowest host flux densities.
For F110W, the GALFIT host flux density is less than a factor of two smaller than our flux density, but given the uncertainties (on both the GALFIT subtraction and our approach), is entirely consistent.
F160W provides a less constrained fit to the host from GALFIT, but again is consistent within the uncertainties.
Here, the flux density is nearly a factor of five lower, but the uncertainty is larger and nearly encompasses the full range of host flux densities from our initial PSF subtraction. 
Thus, more sophisticated subtraction routines do not make a difference for the relevant photometry.
They do, however, introduce more free parameters that we cannot adequately constrain with the available data and so we proceed, albeit cautiously, with our weak detection of the host, but clear detection of the companion based off of our simplistic empirical PSF subtractions.

\subsection{HSC-SSP Photometry}\label{sec:hsc}

\begin{figure*}
    \centering
    \includegraphics[width=0.8\linewidth]{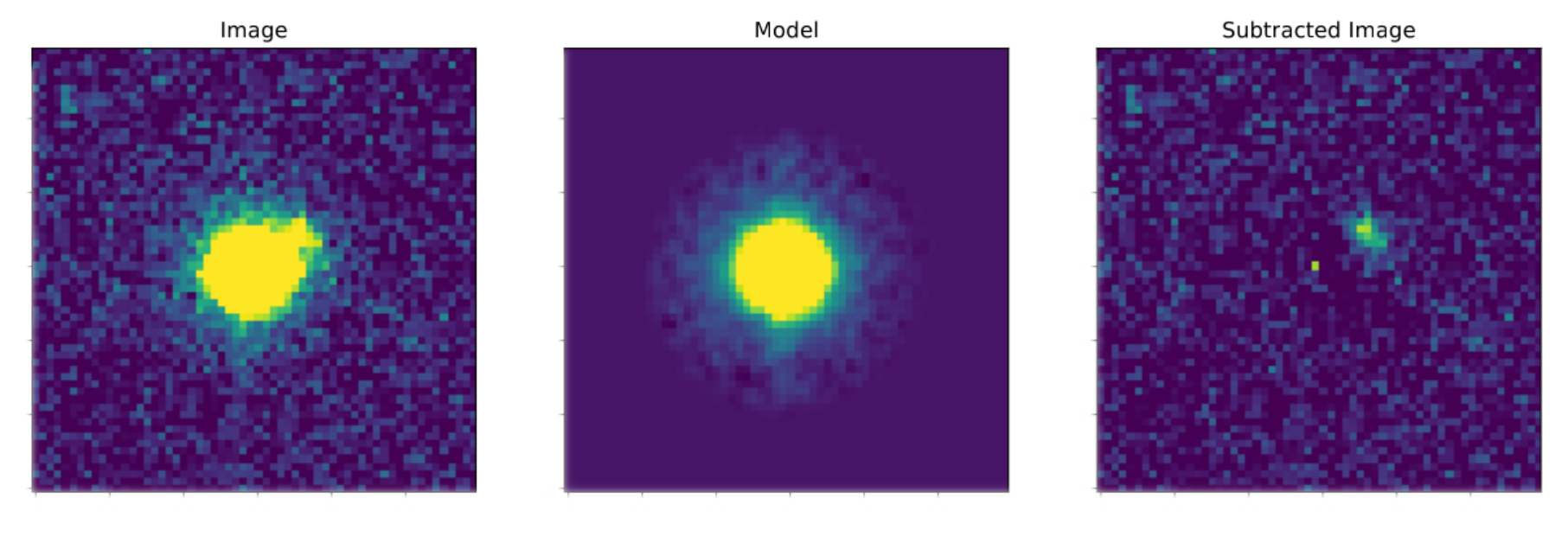}
    \caption{An example of the HSC PSF-subtraction in the \textit{y}-band. Each image is $60 \times 60$ pixels. Only the companion galaxy remains in the residual image.}
    \label{fig:hsc_sub}
\end{figure*}

In addition to the WFC3 data, we extract optical/NIR ancillary data for J1607 from the Hyper Suprime-Cam Subaru Strategic Program (HSC-SSP) Public Data Release 1 (PDR1: \citealt{aihara18b}). 
The ELAIS-N1 field is covered as part of the HSC-SSP Deep layer and includes photometry in the \textit{g}, \textit{r}, \textit{i}, \textit{z}, and \textit{y} bands, reaching limiting magnitudes of 26.8, 26.6, 26.5, 25.6, and 24.8, respectively. 
Similar to the HST observations, due to the close proximity of the quasar host galaxy and the star formation companion, we must subtract the emission from the quasar in order to obtain photometry on the companion. 
Although the size of the PSF does change slightly between each of the HSC bands, it ranges from about 0.3\arcsec - 0.4\arcsec, so we can easily separate the companion from the central quasar.
We use GALFIT to complete the PSF subtraction of the quasar, first using model PSFs produced as part of the HSC-SSP PDR2 \citep{aihara19} and, as a comparison, by fitting a Sersic model with GALFIT. 
We find that the two models both produce good and comparable PSF subtractions except in the \textit{i}-band, where subtracting the HSC-SSP PSF model leaves significant residuals towards the center of the quasar compared to the Sersic model. 
The difference in photometry estimated from the two sets of PSF-subtracted images is negligible. 
An example of the HSC PSF-subtraction is given in Fig. \ref{fig:hsc_sub}.

We estimate the photometry of the offset companion in each of the HSC bands using an elliptical aperture centered on the companion with the semi-major axis aligned with its elongated axis. 
We then increase the size of the semi-minor axis of the aperture and record the aperture sum. 
We find that, for each band, the aperture sum increases as the semi-minor axis increases to include more of the flux from the region of the companion, until the region in which the emission from the quasar host has been subtracted is reached. 
In these regions, the PSF subtraction typically leaves small, negative residuals, and so the aperture flux starts to decrease. 
We therefore use the aperture sum from the semi-minor axis which corresponds to the maximum aperture sum, as this aperture should contain the maximum flux from the region of the companion without contaminating this flux with negative residuals from the PSF subtraction. 
We additionally visually inspect the resulting apertures to ensure proper placement.
We then convert the aperture sum value to an AB magnitude using the standard conversion factors in the HSC-SSP headers. 
In order to estimate the errors on our magnitudes in each band, we combine the rms noise in the image, the Poissonian error on the aperture sum (which measures counts per second), and the standard 1 per cent calibration error \citep{aihara18a} in quadrature, before converting into an AB magnitude.

Though detected in the \textit{r}, \textit{i}, \textit{z}, and \textit{y} bands, the emission from the companion is undetected in the \textit{g} band. 
In order to obtain upper limits on the \textit{g}-band magnitude, we place another elliptical aperture in the same region as in the other bands, but with a semi-minor axis equal to the average of the semi-minor axes of each of the apertures used for the other bands (the semi-minor axis sizes in the different bands are all typically within $\sim 1$ pixel of each other, so the average value should be comparable). 
We then extract the aperture sum in this region and convert this into a \textit{g}-band magnitude.

\subsection{Stellar Mass and SFR} \label{sec:cigale_fits}

Once we have our photometric data for the host galaxy and the companion, we determine the best-fit spectral energy distribution (SED) models using CIGALE \citep[Code Investigating GALaxy Emission;][]{burgarella05, noll09, boquien19, yang22}. 
CIGALE fits galaxy SEDs from the far-UV to the radio and estimates their physical properties. 
For purposes of describing the quasar host and the companion, we focus on two such properties: SFR and stellar mass.

Given the lack of photometry available for both the host and the companion individually, we first use CIGALE to describe the total system (i.e. the quasar, its host, and the companion). 
All of the flux densities used in the fit are provided in Tab. \ref{tab:fluxes}. 
We assume a delayed star formation history with an optional constant burst.
We allow the main stellar population age to vary between 250 - 1400\,Myr with the lower bound chosen to be consistent with the highest-redshift galaxies currently observed by JWST \citep{harikane25}.
The age of the burst is allowed values between 10 and 50\,Myr. 
We additionally use the stellar population synthesis model of \citet{bruzual03} [{\footnotesize \texttt{bc03}}] with the initial mass function of \citet{chabrier03} and a solar metallicity.

\begin{deluxetable}{ccc}
    \tablecaption{Flux Densities Used in SED Fitting}
    \label{tab:fluxes}
    \tablehead{
        \colhead{Telescope/Survey} & \colhead{Wavelength} & \colhead{Flux Density}\\[-0.25cm]
        \colhead{} & \colhead{($\rm \mu m$)} & \colhead{($\rm \mu Jy/mJy$)}
    }
    \startdata
        \multirow{2}{*}{CFHT$\rm ^a$} & 0.4 (\textit{u}) & $4.44 \pm 0.03$ \\
        & 0.9 (\textit{z}) & $208.36 \pm 0.10$\\
        \hline
        \multirow{9}{*}{HSC-SSP$\rm ^{a,b}$} & 0.5 (\textit{g}) & $60.31 \pm 0.03$\\
            && $\boldsymbol{\it0.56^\mathrm{d}}$\\
            & 0.6 (\textit{r}) & $181.82 \pm 0.08$\\
            && $\boldsymbol{\it2.08 \pm 0.02}$\\
            & 0.8 (\textit{i}) & $162.03 \pm 0.07$\\
            && $\boldsymbol{\it2.64 \pm 0.03}$\\
            & 0.9 (\textit{z}) & $\boldsymbol{\it2.23 \pm 0.02}$\\
            & 1.0 (\textit{y}) & $207.27 \pm 0.22$\\
            && $\boldsymbol{\it2.28 \pm 0.03}$\\
            \hline
            \multirow{2}{*}{HST} & 1.1 (F110W) & $244.9 \pm 7.3$\\
            && $\boldsymbol{\it3.1 \pm 0.1}$\\
            & 1.6 (F160W) & $ 287.1 \pm 8.5$\\
            && $\boldsymbol{\it3.3 \pm 0.1}$\\
            \hline
            UKIDSS$\rm ^a$ & 2.2 (\textit{K}) & $365.6 \pm 1.4$\\
            \hline
            \multirow{4}{*}{IRAC$\rm ^a$} & 3.6 & $356.8 \pm 0.7$\\
            & 4.5 & $416.1 \pm 1.3$\\
            & 5.8 & $661.8 \pm 12.8$\\
            & 8.0 & $1337.1 \pm 4.6$\\
            \hline
            \multirow{2}{*}{MIPS} & 24$\rm ^a$ & $5.7 \pm 1.1$$\rm ^e$\\
            & 70$\rm ^c$ & $22.2 \pm 4.4$$\rm ^e$\\
            \hline
            \multirow{2}{*}{PACS$\rm ^a$} & 100 & $58.0 \pm 17.8$\\
            & 160 & $44.4 \pm 26.0$\\
            \hline
            \multirow{3}{*}{SPIRE$\rm ^a$} & 250 & $65.3 \pm 1.9$\\
            & 350 & $55.1 \pm 2.9$\\
            & 500 & $28.6 \pm 7.1$\\
            \hline
            SCUBA$\rm ^c$ & 850 & $31.0 \pm 6.2$$\rm ^e$\\
            \hline
            MAMBO$\rm ^c$ & 1200 & $7.3 \pm 0.9$\\
            \hline
            SMA & 1300 (230\,GHz) & $\boldsymbol{\it 3.13 \pm 0.36}$\\
    \enddata
    \tablecomments{The values in bold italics are used to describe the sub-mm companion, while the rest describe the total system (quasar + host + companion). The first part of the table (up to and including the IRAC data) provides the flux densities in $\mu$Jy, while the second part gives them in mJy.\\
    $\rm ^a$ Taken from the HELP catalog \citep{shirley19}.\\
    $\rm ^b$ See Sec. \ref{sec:hsc} for info on extracting the companion's flux densities.\\
    $\rm ^c$ \cite{clements09} and references therein. \\
    $\rm ^d$ This represents an upper limit.\\
    $\rm ^e$ To be conservative, we adopted a 20\% uncertainty instead of the catalog value.
    }
\end{deluxetable}

To describe the AGN, we use the smooth torus models of \cite{fritz06}. 
We also fit the AGN component using the SKIRTOR \citep{stalevski12,stalevski16} module, which describes a clumpy two-phase torus. 
The resultant AGN luminosity is lower for the clumpy torus, but by less than a factor of two. 
All other values from the SED fits were consistent with each other within the errors. 
We therefore proceed with the fit assuming the smooth torus, but our results would be largely unchanged if we instead assumed a clumpy torus. 

Lastly, we rely on the models of \cite{dale14} to describe the dust emission. 
As with the AGN component, we also tested the other dust emission modules \citep[e.g.][]{casey12, draine14}. 
We found that they each produced reasonable fits to the data, as evidenced by their (reduced) chi-squared values. 
Though the \cite{draine14} module has more parameters, it does not significantly improve the fit, so we adopt the simpler \cite{dale14} module.
Here, the AGN fraction is set to zero since we have included a separate AGN component.

\begin{figure*}
    \centering
    \includegraphics[width=0.7\linewidth]{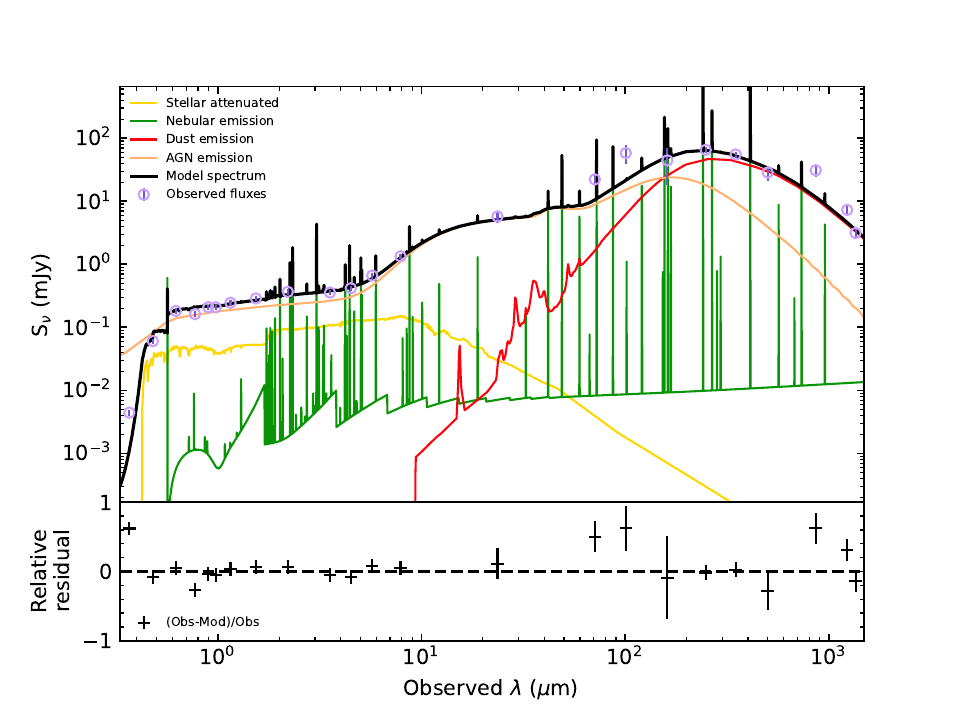}
    \caption{SED fit for the total system. The total fit is depicted by the solid black curve. The orange line describes the AGN component, while the red line shows the starburst component. For a comparison of the stellar fluxes between the CIGALE fit, our PSF subtractions, and the GALFIT PSF subtractions, see \S \ref{sec:hst_phot} and \S \ref{sec:cigale_fits}.}
    \label{fig:sed_tot}
\end{figure*}

The allowed parameter values for each of the modules defined above are provided in Tab. \ref{tab:cigale_param}. 
The resultant SED fit is given in Fig. \ref{fig:sed_tot}. 
This gives an SFR (averaged over the last 10\,Myr) of $(4500 \pm 500)$\sfr. 
If we were to consider the SFR averaged over the last 100\,Myr, the total system would instead show an SFR of ($2200 \pm 900$)\sfr. 
However, the burst age from the SED fit is roughly 30\,Myr, so we proceed with the 10\,Myr SFR rather than the 100\,Myr one as it better describes the current state of the system.
The fit further gives a stellar mass of $(5.8 \pm 3.0) \times 10^{11}$\,M$_\odot$. 

\begin{deluxetable*}{lc}
    \tablecaption{Allowed Model Parameters in Total System CIGALE Fit to the SED}
    \label{tab:cigale_param}
    \tablehead{
        \colhead{Parameter} & \colhead{Value}\\[-0.5cm] 
    }
    \startdata
    \multicolumn{2}{c}{Star formation history [{\scriptsize \texttt{sfhdelayedbq}}]} \\
    \hline
    main stellar population e-folding time [Myr] & 100, 500, 1000 \\
        main stellar population age [Myr] & 250 - 1400; increments of 50 \\
        burst age [Myr] & 10, 20, 30, 40, 50 \\
        ratio of SFR after/before & 0.1, 0.5, 1, 2, 5, 10, 25 \\
        \hline
        \multicolumn{2}{c}{Dust attenuation [{\scriptsize \texttt{dustatt\_modified\_starburst}}]}\\
        \hline
        amplitude of the UV bump & 3.0\\
        \hline
        \multicolumn{2}{c}{Dust emission [{\scriptsize \texttt{dale2014}}]}\\
        \hline
        AGN fraction & 0.0 \\
        alpha slope & 0.0625, 1.0, 2.0, 3.0, 4.0 \\
        \hline
        \multicolumn{2}{c}{AGN [{\scriptsize \texttt{fritz2006}}]}\\
        \hline
        torus outer-to-inner radii ratio & 60\\
        optical depth at 9.7\,\mum & 6.0\\
        torus dust exponent for density gradient along radial direction & -1.0, -0.5, 0.0\\
        torus dust exponent for density gradient along polar direction & 0.0, 6.0\\
        opening angle of torus [deg] & 60, 100, 140\\
        angle between equatorial axis and line-of-sight & 89.99\\
        AGN fraction & 0.0, 0.1, 0.3, 0.5, 0.7, 0.9, 0.99 \\
    \enddata
    \tablecomments{The angle between the equatorial axis and line-of-sight is chosen as J1607 is a type 1 quasar. The same parameters are used when fitting the companion, though the AGN component is no longer considered.}
\end{deluxetable*}

When looking at the probability distribution function (PDF) of the main stellar population age, it has a double-peaked profile at smaller values and a local maximum at higher values.
Thus, to test the effects of the main population age on the SFR and stellar mass, we run two additional fits: one where the age is allowed to vary between 250 - 750\,Myr (i.e. the double-peaked part of the PDF) and one where the age varies between 800 - 1400\,Myr (i.e. the increase in the PDF to higher values).
The first run, equivalent to a younger system, gives an SFR of $4400\pm600$\sfr and a stellar mass of $(5.5\pm3.2)\times10^{11}$\,M$_\odot$.
An older system gives an SFR of $4700\pm400$\sfr and a stellar mass of $(6.1\pm2.8)\times10^{11}$\,M$_\odot$.
Given the uncertainties, each of the fits are consistent with one another, so we proceed with the fit covering the full range of ages.

We next consider the CIGALE-predicted attenuated stellar flux densities in our two HST filters and compare them to our empirical-PSF-subtracted values (here the sum of the host and companion values in Table \ref{tab:sub_comp}).
The CIGALE flux densities end up being a factor of 2-3 higher than expected, but the uncertainties on our flux densities are large enough that the two are just consistent within $2.3\sigma$ for F110W and $1.7\sigma$ for F160W.

\begin{figure*}
    \includegraphics[width=0.3\linewidth]{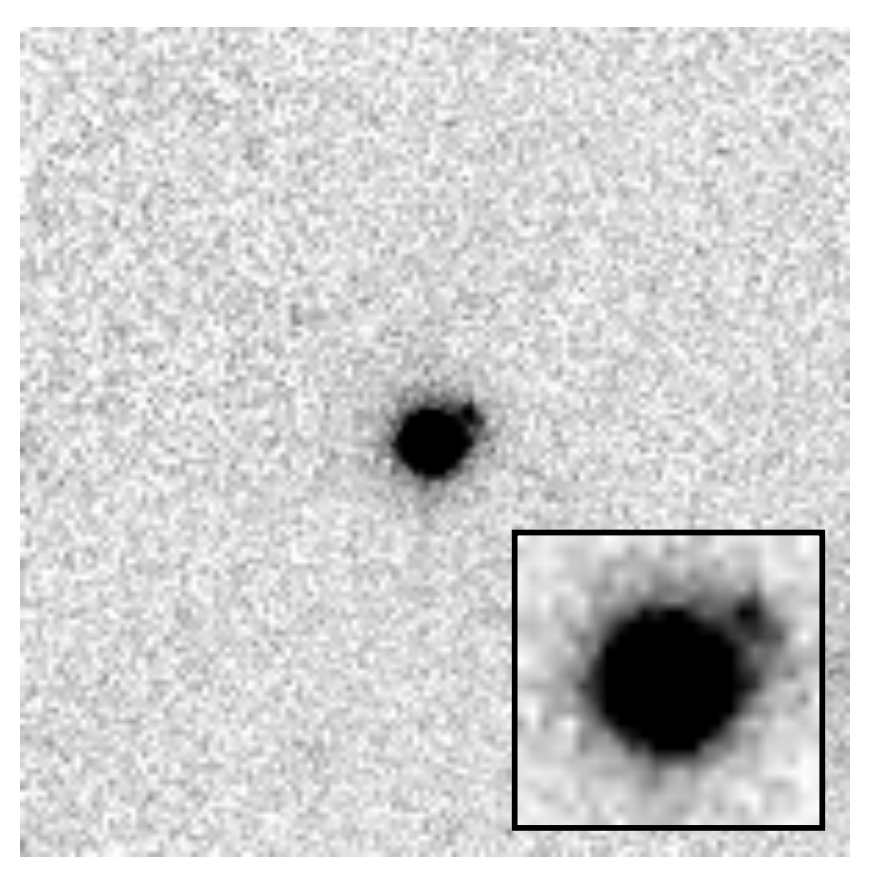}
    \includegraphics[width=0.3\linewidth]{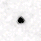}
    \includegraphics[width=0.3\linewidth]{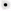}
    \caption{J1607 images. From left to right: HSC-SSP \textit{y}, IRAC 3.6$\mu$m, and MIPS 24$\mu$m. All images are 25" on a side with north up and east to the left. Within these cutouts, the companion is only visible in the HSC-SSP image, as seen by the zoom-in of the inner 2\arcsec x 2\arcsec.}
    \label{fig:cutouts}
\end{figure*}

Though we do not have enough data available to fully constrain the respective SEDs of either the companion or the host, we do attempt to describe the former using what is available. 
Because the companion does not contain a bright central source, it has more photometry than does the host; we have both the WFC3 and the HSC-SSP data. 
Several cutouts are provided in Fig. \ref{fig:cutouts}, which show that we are unable to separate the quasar host and companion in all images other than those from HSC-SSP (and of course the HST images in Fig. \ref{fig:psf_sub_comp}).

We additionally include data from new observations with the Submillimeter Array \citep[SMA;][]{ho04} at 230\,GHz, corresponding to 1.3\,mm. 
As older sub-mm observations showed the peak of the sub-mm emission to come from the companion rather than the host \citep{clements09}, we attribute all of the sub-mm flux to it. 
This is not to say that the host galaxy of the quasar has no dust; it just implies that the host has less dust than does the companion. 
We leave a more detailed description of the SMA observations to future work (Cairns et al., in prep). 

\begin{figure}
    \centering
    \includegraphics[width=\linewidth]{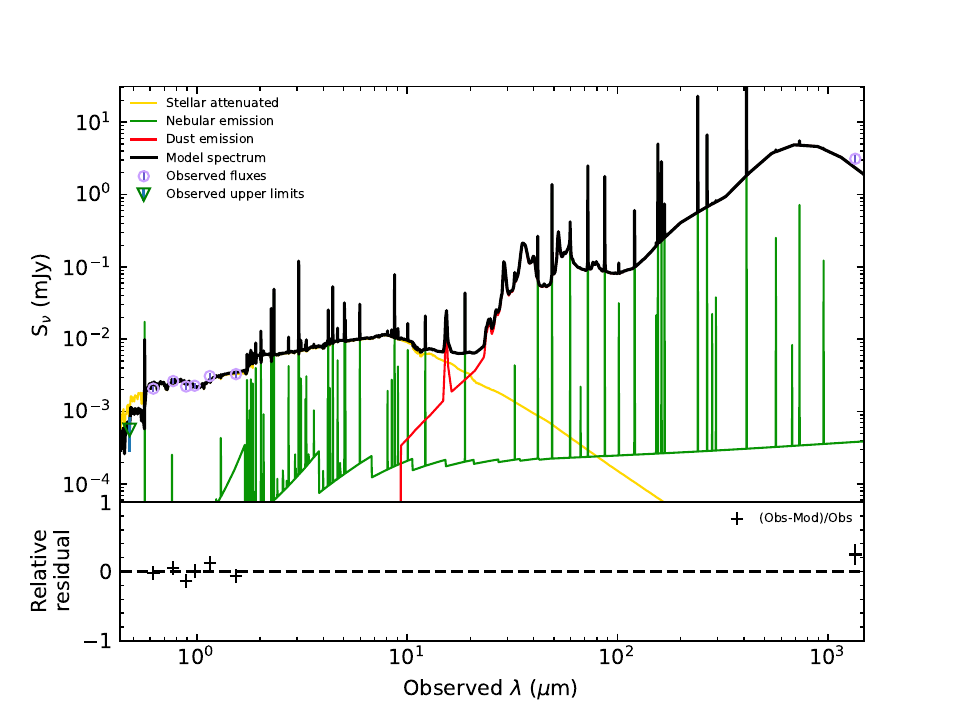}
    \caption{SED fit for the offset star formation companion. The total fit is depicted by the solid black curve, while the starburst component used to describe the sub-mm data is shown by the red line.}
    \label{fig:sed_blob}
\end{figure}

In fitting the SED, we use the same parameters as those provided in Tab. \ref{tab:cigale_param}, neglecting the AGN component. 
The best-fit to the SED for the companion is shown in Fig. \ref{fig:sed_blob}. 
From the fit, we estimate an SFR (again averaged over the last 10\,Myr) of ($180 \pm 40$)\sfr and a stellar mass of $(7.9 \pm 5.0) \times 10^{10}$\,M$_\odot$.

Since we only have the two HST points for the host, CIGALE does not prove useful in describing the SFR or stellar mass of the galaxy. 
We can, however, roughly estimate these values as our total system is comprised of the quasar/its host galaxy and its companion. 
Thus, the difference (in e.g. SFRs) between the total system and the companion gives rough estimates of the host's properties. 
This gives an SFR of ($4300 \pm 500$)\sfr and a stellar mass of $(5.0 \pm 3.1) \times 10^{11}$\,M$_\odot$. 
The SFRs and stellar masses are summarized in Table \ref{tab:results}.

\begin{table}
    \begin{center}
        \caption{System Properties from CIGALE SED Fits} 
        \label{tab:results}
        \begin{tabular}{lc}
            \hline
            \hline
            \multicolumn{2}{c}{total system}\\
            \hline
            SFR [\sfr] & $4500 \pm 500$ \\
            $M_*$ [$10^{11}$\,M$_\odot$] & $5.8 \pm 3.0$\\
            \hline
            \multicolumn{2}{c}{companion} \\
            \hline
            SFR [\sfr] & $180 \pm 40$ \\
            $M_*$ [$10^{10}$\,M$_\odot$] & $7.9 \pm 5.0$\\
            \hline
            \multicolumn{2}{c}{host} \\
            \hline
            SFR [\sfr] & $4300 \pm 500$ \\
            $M_*$ [$10^{11}$\,M$_\odot$] & $5.0 \pm 3.1$\\
            $L_\mathrm{AGN} [10^{13}\,\mathrm{L}_\odot]$ & $11.6 \pm 0.6$ \\
            M$_\mathrm{BH}$ [10$^9$\,M$_\odot$] & $3.5 \pm 0.2$\\
            \hline
        \end{tabular}
    \end{center}
    \tablecomments{As we do not have enough photometry for the host to constrain its SED, we instead take the difference between the total system properties and the companion properties to describe the host. The SFRs are averaged over the last 10\,Myr.}
\end{table}

There is a possibility that the quasar light is being scattered by the interstellar medium. 
In such a scenario, some of the quasar light would still be contributing to the host flux, even after the PSF has been subtracted. 
As the quasar exhibits broad lines (and therefore is unobscured), we do not expect the host to have a comparatively high amount of obscuration.
If both the AGN and its host are relatively unobscured, the contribution of the AGN in the host residuals is likely negligible. 
As we cannot address this potential issue further, we proceed (albeit cautiously) with the above SFR and stellar mass values for the host given the limited data available.

\subsection{AGN Luminosity}

From the fit to the total system SED, we extract an AGN luminosity of $(11.6 \pm 0.6) \times 10^{13}$\,L$_\odot$.
This is about a factor of two higher than that from \cite{clements09}, but we have incorporated newer observations taken from the Herschel Extragalactic Legacy Project \citep[HELP;][]{shirley19} catalog. 
We additionally fit the fluxes to the Type 1 quasar templates of \citet{polletta07} and again obtain a value within about a factor of two, L$_\textnormal{AGN} \sim 6.1 \times 10^{13}$\,L$_\odot$.

If we assume Eddington-limited accretion, this gives a soft lower limit on the black hole mass of log$(M_\mathrm{BH}\,[\mathrm{M}_\odot]) = 9.55$. 
If we instead adopt the redshift-dependent Eddington ratio (i.e. the fraction of the Eddington limit at which the AGN is actually accreting) of \citet{shan13}, which is 0.41 at $ z = 3.65$, we obtain a black hole mass estimate of log$(M_\mathrm{BH}\,[\mathrm{M}_\odot]) = 9.93 \pm 0.30$. 
This agrees within the uncertainties with the virial black hole mass estimate of \cite{rakshit20}, log$(M_\mathrm{BH}\,[\mathrm{M}_\odot]) = 9.74 \pm 0.08$, which was found using the C{\scriptsize IV} emission properties of J1607 and the \cite{vestergaard06} relation. 

Considering both the black hole mass and the stellar mass of the host, we find that J1607 is consistent with other quasars at similar redshifts \citep{kormendy13}. 
It is additionally comparable to the local elliptical population \citep[e.g.][]{kormendy13, reines15}. 


\section{Discussion}\label{sec:discussion}

J1607 is a case study of an extremely luminous early-stage merger 1.7\,Gyr after the Big Bang. 
It harbors both a rapidly accreting SMBH and two separate star-forming progenitors. 
Among the small number of $z>3$, luminous galaxies that have been studied in detail \citep[e.g.][]{ma15, riechers17, pavesi18, shao19}, very few have these properties \citep[e.g.][]{decarli17}. 
J1607 thus offers a relatively unexplored window into early, luminous galaxy assembly. 
We first discuss the properties of the two galaxies (AGN host and sub-mm companion) in context with other galaxies. We then show one possible future path for the merger. 

\subsection{Comparison with other high-redshift galaxies}

There exists a correlation between SFR and stellar mass among the majority of the star-forming galaxy population  called the star-forming `main sequence' \citep[e.g.][]{noeske07, speagle14}. 
This main sequence evolves with redshift; that is, at higher redshifts, main sequence galaxies exhibit higher SFRs on average compared to their lower redshift counterparts \citep[e.g.][]{elbaz11, rodighiero11, schreiber15}. 
Starbursts lie above this main sequence as they are forming new stars at higher than average rates.
The star formation observed in starbursts is often attributed to major mergers \citep{daddi10}, while main sequence star formation is likely due to internal secular processes. 
This does not, however, preclude main sequence galaxies from being involved in ongoing mergers \citep{sparre15}.

Assuming the main sequence relation (Equation 28) of \cite{speagle14} and the stellar mass resultant from the CIGALE fit, the total J1607 system lies above the main sequence by at least a factor of two.
If we assume that the total system is only the quasar host and the sub-mm companion, the host similarly lies above the main sequence at its assumed stellar mass. 
The companion, however, is consistent with the low end of the main sequence\footnote{If we instead consider the high-redshift, uncalibrated description \citep[Equation 30 of ][]{speagle14}, the uncertainties are large enough such that both the host and the companion are consistent with the main sequence. The host is also consistent with the high-end of the main sequence when considering the SFR averaged over the last 100\,Myr.}.
It therefore is plausible to assume that J1607 is a merger wherein the resultant burst in star formation in the quasar host causes it to lie above the main sequence, consistent with mergers triggering the most extreme starbursts at all epochs.

\subsection{Galaxy assembly through mergers}

Given its redshift and its active assembly of stellar and SMBH mass, J1607 is a candidate for being a major assembly episode of a massive, quiescent galaxy by $z \sim 2$. 
In this section we constrain the properties of the quiescent galaxy that J1607 may signpost the assembly of and how these properties compare with the expected descendants of other far-infrared luminous galaxies.  
To do so, we consider the star formation rate of J1607 in context with its stellar mass and molecular gas mass. 
This requires an estimate of the molecular gas content of J1607 and suitable comparison populations. 

We use two literature studies of the molecular gas in J1607.
\cite{iono12} present a detection of $^{12}$CO \mbox{\textit{J}=4--3}, a tracer of the gas in star-forming regions, with an estimated total gas mass of $M_\textnormal{gas} = (1.1 \pm 0.2) \times 10^{11}\,\textnormal{M}_\odot$. 
\cite{fogasy22} present a detection of the $^{12}$CO \mbox{\textit{J}=1--0} line, which traces cold gas. 
They split the CO emission into two main components: the quasar host and a companion that is $\sim$16.8\,kpc to the northwest of the quasar host. 
The gas masses for the host and the CO companion, respectively, are $(2.4 \pm 0.9) \times 10^{10}\,\textnormal{M}_\odot$ and $(2.6 \pm 1.3) \times 10^{10}\,\textnormal{M}_\odot$. 
We thus consider the following components of J1607:  the system as a whole, the quasar host, the sub-mm companion (located 11\,kpc to the northwest of the host), and the CO companion (located 17\,kpc to the northwest of the host).  
As comparison samples\footnote{Where necessary, all masses have been converted assuming a Chabrier initial mass function for the stellar masses and a CO-to-H$_2$ conversion factor of \mbox{0.8 M$_\odot$ (K\,km\,s$^{-1}$pc$^2$)$^{-1}$} for the molecular gas masses.}, we consider the following. 
First, we compile other hyperluminous infrared galaxies at high redshift with similar SFRs to J1607 \citep{riechers13, riechers17, nayy17, marrone18, pavesi18}. 
Second, we include a collection of sub-mm bright galaxies (SMGs) at high redshift, which are infrared-luminous but typically have SFRs about a factor of five lower than in J1607 \citep{engel10, bothwell13, fu13}.  
Third is a sample of $z\sim0$ ULIRGs \citep{farrah22}. 
Though they are about an order of magnitude less luminous than J1607, these ULIRGs are the most luminous sources in the nearby universe and serve as a baseline from which to estimate possible trends with redshift.  
Finally, to extend the comparison with $z\sim0$ systems to lower luminosities, we include local AGN \citep{husemann22, smirnova22} and normal galaxies \citep{saintonge17}, both of which are about an order of magnitude less luminous than the local ULIRG sample. 

\begin{figure}
    \centering
    \includegraphics[width=1.0\linewidth]{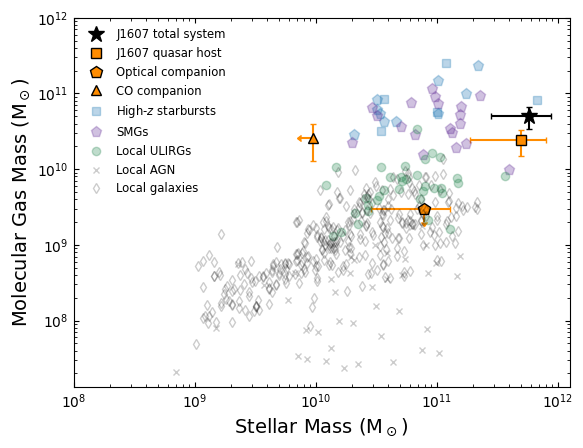}
    \caption{Molecular gas vs stellar mass. We consider the total J1607 system (black star), as well as each of its components (shown in orange). We then compare to high-redshift starbursts \citep[blue squares;][]{riechers13, riechers17, nayy17, marrone18, pavesi18}, high-redshift SMGs \citep[purple pentagons;][]{engel10, bothwell13, fu13}, local ULIRGs \citep[green circles;][]{farrah22}, local AGN \citep[black crosses;][]{husemann22, smirnova22}, and the normal local galaxy population \citep[black diamonds;][]{saintonge17}. The quasar host aligns with other high-redshift systems. Since the CO companion likely has a lower stellar mass than reported here, it is extremely gas-rich. The sub-mm companion is comparatively gas-poor, perhaps due to gas-stripping over the course of the merger. }
    \label{fig:mgas_mstar_comp}
\end{figure}

The three components of J1607, as well as the comparison samples, are shown in the $M_\mathrm{gas}-M_{*}$ plane in Figure \ref{fig:mgas_mstar_comp}. 
Considering first J1607: the components of this system have an interesting spread in properties.  
The quasar host has a stellar mass of $(5.0\pm3.1)\times10^{11}$M$_{\odot}$ and a gas mass of $(2.4\pm0.9)\times10^{10}$\sol, giving a gas mass fraction of $\sim0.05$. 
The sub-mm companion is detected by HST but is not detected in CO (with the caveat that about ten percent of the \citet{fogasy22} CO emission overlaps spatially with the sub-mm emission). 
This gives a stellar mass of $(7.9\pm5.0)\times10^{10}$M$_{\odot}$ and an upper limit on the molecular gas mass of $\lesssim3\times10^{9}$\sol.
Its gas mass fraction is thus at most 0.04 (0.09 when considering the uncertainty on the stellar mass). 
In contrast, the CO companion is not detected in the HST imaging.
It has a molecular gas mass of  $(2.6\pm1.3)\times10^{10}$M$_{\odot}$ and an upper limit on the stellar mass of $\lesssim9.6\times10^{9}$M$_{\odot}$.
This makes the CO companion's gas mass fraction at least 0.73 (0.58 when considering the large uncertainty on the gas mass), illustrating the diversity in gas mass fractions that is possible within components of a galaxy at high redshift and highlighting the dynamically complex nature of high-redshift active galaxies. 

Turning to comparisons between J1607 and other populations:  J1607 as a whole aligns with the comparison hyperluminous starburst and SMG samples, though on the higher end in terms of stellar mass.  
The quasar host is also consistent with the other hyperluminous starbursts and SMGs. 
The sub-mm and CO emission, however, are not. 
The gas-poor sub-mm companion is in the same part of the $M_\mathrm{gas}-M_*$ parameter space as the \textit{local} ULIRG population, possibly even lower.  
The gas-rich CO companion is on the edge of the distribution of other high-redshift infrared-luminous galaxies, making it one of the most gas-rich objects known at high redshift.  
Neither the sub-mm companion nor the CO companion show any evidence for AGN activity (though we cannot rule it out). 
We therefore argue that the sub-mm companion is gas-poor because molecular gas has been tidally stripped from it due to the dynamics of the merger.
This lends support to the idea of merger-induced gas stripping as a mechanism distinct from AGN feedback, to quench star formation in high-redshift galaxies. 
A plausible scenario is one in which a companion galaxy infalling towards the quasar host had its gas largely stripped from it during infall, leading to the now spatially separated sub-mm companion and CO companion.  
Improved constraints on the velocity field of the gas and stars in J1607 would however be required to confirm or refute this idea. 

We now consider possible futures for J1607. 
If all of the detected gas reservoir in J1607 acts as fuel for star formation with 100 percent efficiency during the current burst, then the depletion time for the gas, assuming the \cite{iono12} mass, is ($24 \pm 5$)\,Myr.
This is smaller than the depletion timescales for most SMGs \citep[e.g.][]{greve05, riechers11}, but does align with other rapidly star-forming quasars \citep[e.g.][]{riechers06}.  
The resulting total stellar mass is \mbox{$(4 - 10) \times 10^{11}\,\mathrm{M}_\odot$}.  
This is comparable to the total stellar masses of massive early-type galaxies at $z\sim0$ \citep[e.g.][]{ma14, mehrgan19}.
Such systems also exist at $z\sim2$, in small numbers \citep[e.g.][]{cimatti04, daddi05, damjanov09, sherman20}. 
Considering instead the components of J1607, and assuming the CO companion does not merge, gives a similar result. 
Assuming the quasar host converts 100 percent of its gas to stars, then its depletion time is $(6 \pm 2)$\,Myr and its final total stellar mass is \mbox{$(2 - 8) \times 10^{11}\,\mathrm{M}_\odot$}. 
A similar analysis for the sub-mm companion yields a final total stellar mass of \mbox{$(3 - 13) \times 10^{10}\,\mathrm{M}_\odot$}. 
If the host and the sub-mm companion then merge, the resultant system would have a stellar mass in the range of \mbox{$(2 - 9) \times 10^{11}\,\mathrm{M}_\odot$}. 
The consistency between the predicted final total stellar mass of J1607 and massive local early-type galaxies, along with the likely endpoints of major mergers as early-type systems, supports the idea that at least some massive ETGs locally complete the bulk of their assembly by $z\sim2$. 

\section{Conclusions}\label{sec:conclusions}

We have presented HST WFC3 imaging data in F110W and F160W for J1607 at $z = 3.65$. 
We show that this system is comprised of three components; an AGN host galaxy, and two companion sources at projected separations of 11kpc and 17kpc, in the same direction. 
To investigate the properties of this system, we first remove the light from the AGN via a careful subtraction of the quasar PSF, and then combine the HST imaging with archival data on the molecular gas content of the system. 
We test two methods for creating the PSFs, one built from a nearby star and one that is synthetic, and find that the former is better able to reproduce the diffraction spikes. 
We extract photometry for all three sources, and combine with archival data to fit the the SEDs with CIGALE to determine their physical properties. 
The AGN host SED is consistent with a stellar mass of $(5.0 \pm 3.1) \times 10^{11}$M$_\odot$ and an SFR of ($4300 \pm 500$)\sfr. 
The closer companion SED gives a stellar mass of $(7.9 \pm 5.0) \times 10^{10}$M$_\odot$ and an SFR of ($180 \pm 40$)\sfr.
The AGN host is consistent with a very massive, extremely star-forming galaxy lying above the stellar mass - star formation rate main sequence at this epoch, while the closer companion likely lies on the main sequence. 
The SMBH mass of the AGN host, assuming Eddington-limited accretion, is $\sim 3.5 \times 10^{9}$M$_\odot$. 
The total stellar mass-to-SMBH mass ratio of the AGN host is consistent with the locally observed relation for massive early-type galaxies. Regardless of whether the closer companion merges with the AGN host, and how long their star formation episodes last, these properties mark J1607 as being the primal assembly episode of a massive early-type galaxy. 

The two companion sources present markedly different stellar and molecular gas properties. 
The closer companion is not detected in CO, giving an upper limit on its molecular gas mass of $\lesssim3\times10^{9}$\sol. 
The more distant companion is clearly detected in CO, with a molecular gas mass of $(2.6\pm1.3)\times10^{10}$M$_{\odot}$, but is not detected in our HST imaging, setting an upper limit on its stellar mass of $\lesssim 9.6\times10^{9}$M$_{\odot}$. 
Since neither companion shows evidence for AGN activity, and both lie on the same radial direction, we propose that this may signpost a form of merger-driven quenching. 
In this scenario, the dynamics of the merger, as well as or instead of AGN activity, are responsible for stripping molecular gas from the closer companion and thus inhibiting its star formation. 
This highlights the possibility that merger-driven feedback could play an important role in quenching star formation at high redshift.

\section*{Acknowledgements}

We thank the referee for a helpful report.
This research is based on observations made with the NASA/ESA Hubble Space Telescope obtained from the Space Telescope Science Institute, which is operated by the Association of Universities for Research in Astronomy, Inc., under NASA contract NAS 5–26555. 
These observations are associated with program 15200. 
The Submillimeter Array is a joint project between the Smithsonian Astrophysical Observatory and the Academia Sinica Institute of Astronomy and Astrophysics and is funded by the Smithsonian Institution and the Academia Sinica.
The Herschel Extragalactic Legacy Project, (HELP), is a European Commission Research Executive Agency funded project under the SP1-Cooperation, Collaborative project, Small or medium-scale focused research project, FP7-SPACE-2013-1 scheme, Grant Agreement Number 607254. DLC, and JC acknowledge support from STFC, in part through grant numbers ST/K001051/1 and ST/V005359/1.

For the purpose of open access, the author(s) has applied a Creative Commons Attribution (CC BY) license to any Author Accepted Manuscript version arising.






\bibliography{j1607_hst}{}
\bibliographystyle{aasjournal}

\end{document}